\documentclass[12pt]{article}
\usepackage{amssymb,amsmath,amsbsy,amsthm}
\usepackage{latexsym,graphicx}
\usepackage{authblk}

\theoremstyle{plain}
\newtheorem{theorem}{Theorem}[section]
\newtheorem{lemma}[theorem]{Lemma}

\theoremstyle{definition}
\newtheorem{definition}[theorem]{Definition}
\theoremstyle{remark}

\newcommand{\Ref}[1]{(\ref{#1})}

\renewcommand\Authfont{\scshape\normalsize}
\renewcommand\Affilfont{\itshape\small}
\author[1,*]{Jonas de Woul}
\author[2,\dag]{Jens Hoppe}
\author[2,\ddag]{Douglas Lundholm}
\affil[1]{Department of Theoretical Physics, Royal Institute of Technology (KTH)\newline SE-106 91 Stockholm, Sweden \vspace{2mm}}
\affil[2]{Department of Mathematics, Royal Institute of Technology (KTH)\newline SE-100 44 Stockholm, Sweden \vspace{2mm}}
 
\title{\Large{\bf{Partial Hamiltonian reduction of relativistic extended objects in light-cone gauge}}}

\date{\vspace{-0.5cm}\small\today\vspace{-0.5cm}}

\begin{document}

\maketitle

\let\oldthefootnote\thefootnote
\renewcommand{\thefootnote}{\fnsymbol{footnote}}
\footnotetext[1]{{\,jodw02@kth.se}}
\footnotetext[2]{{\,hoppe@math.kth.se}}
\footnotetext[3]{{\,dogge@math.kth.se}}
\let\thefootnote\oldthefootnote

\vspace{-0.5cm}

\begin{abstract}
The elimination of the non-transversal field in the standard light-cone formulation of higher-dimensional extended objects is formulated as a Hamiltonian reduction.
\end{abstract}

\section{Introduction} 
While a certain, partially gauge fixed, light-cone formulation of higher dimensional extended objects has, for quite some time \cite{Hoppe1982}, been known to yield a \emph{polynomial} Hamiltonian (density), with the resulting field equations easily checkable by comparing with the corresponding (gauge-fixed) Lagrangian equation of motions, the final step, the elimination of the dynamical longitudinal field and its canonically conjugate momentum (up to their overall integrals) remained --- including most string-theory reviews --- somewhat unclear concerning the reduction of the Poisson-structure (despite of \cite{HansonReggeTeitelboim1976}, in which the string case is treated, and \cite{BergshoeffSezginTanii1988}, in which a Hamiltonian formulation of the supermembrane is given, --- as well as \cite{Hoppe2002}). As that reduction to (almost) purely transversal degrees of freedom recently also turned out to be related to generalisations of the Witt-Virasoro algebra, and a novel dynamical symmetry \cite{Hoppe2010a,dWHLS,Hoppe2010b,Hoppe2010c}, it seems a good idea to give a detailed account of the Hamiltonian reduction (cp. Theorem \ref{Reduction Theorem}) corresponding to the elimination of the longitudinal field.

\section{Preliminaries}
In this section, we recall the Hamiltonian formulation of the extremal volume problem for relativistic extended objects in light-cone gauge. 

Consider a $M+1$-dimensional submanifold $\mathcal{M}$ in $D$-dimensional Minkowski space with metric $\eta$ of signature $(+,-,\ldots,-)$. Requiring that the Dirac-Nambu-Goto action (volume functional)
\begin{equation}
\label{Dirac-Nambu-Goto action}
S := \int \mathcal{L}\, d\varphi^0 d^M\varphi 
\end{equation}
\begin{equation}
\mathcal{L}:=-\sqrt{\mathcal{G}}, \quad \mathcal{G}:=|\det(\mathcal{G}_{\alpha\beta})|, \quad \mathcal{G}_{\alpha\beta} := \partial_\alpha x^\mu \partial_\beta x^\nu\eta_{\mu\nu}
\end{equation}
is stationary under variations of the embedding functions $x^\mu$, $\mu = 0,1,\ldots D-1$, gives the equations of motion
\begin{equation}
\label{Equations of motion}
\frac{1}{\sqrt{\mathcal{G}}}\partial_\alpha (\sqrt{\mathcal{G}}\mathcal{G}^{\alpha\beta}\partial_\beta x^\mu) = 0
\end{equation}
where $\varphi^0,\varphi^1,\ldots,\varphi^M$, collectively denoted by $(\varphi^0,\varphi)$, are local coordinates on $\mathcal{M}$, and $\partial_\alpha := \frac{\partial}{\partial \varphi^\alpha}$.

For a Hamiltonian formulation, one introduces canonical momenta\footnote{Derivatives with respect to time $\varphi^0$ are written $\dot f:=\partial_0 f $.}
\begin{equation}
\label{Canonical momenta}
p_\mu := \frac{\partial \mathcal{L}}{\partial \dot{x}^\mu}, \quad \mu=0,\ldots,D-1
\end{equation}
satisfying the dynamical Poisson brackets\footnote{The equal time coordinate $\varphi^0$ is suppressed here and in the following.} (all other brackets zero)
\begin{equation}
\label{Canonical Poisson brackets}
\left\{ {x^\mu(\varphi),p_\nu(\tilde\varphi)} \right\}_{PB} = \delta_\nu^\mu\delta(\varphi,\tilde\varphi).
\end{equation}
Because of the general reparameterisation invariance of the action \Ref{Dirac-Nambu-Goto action}, the definition \Ref{Canonical momenta} leads to first class primary 
constraints\footnote{See for example \cite{HansonReggeTeitelboim1976,Dirac1964} for an explanation of the terminology.} (see also \cite{Moncrief2006})
\begin{equation}
\begin{split}
\label{All primary constraints}
C_0  &:= \frac12\left(p_\mu p^\mu - g \right) \approx 0,\\
C_a  &:= p_\mu \partial_a x^\mu \approx 0, \quad a=1,\ldots,M,
\end{split}
\end{equation}
with
\begin{equation}
g := \left|\det\left({\left. \mathcal{G}_{ab}\right|_{a,b=1,\ldots,M}}\right)\right|.
\end{equation}
After the partial gauge fixing
\begin{equation}
\Phi_0 := \frac{x^0+x^{D-1}}{2} - \varphi^0 \approx 0,
\end{equation}
the constraint $C_0$ becomes second class,
\begin{equation}
\left\{ {\Phi_0 (\varphi ),C_0 (\tilde \varphi )} \right\}_{PB} = \pi(\varphi ) \delta (\varphi, \tilde \varphi ),
\end{equation}
with the canonical momenta conjugate to $\zeta := x^0-x^{D-1}$ defined as
\begin{equation}
\pi:=\frac{\partial \mathcal{L}}{\partial \dot{\zeta}},
\end{equation}
while the constraints $C_a$, $a=1,\ldots,M$, remain first class. 
Following \cite{Dirac1964}, one introduces a corresponding ``Dirac bracket'' (assuming $\pi\neq 0$)
\begin{equation}
\begin{split}
\left\{ {F_1 ,F_2 } \right\}_{DB} :=& \left\{ {F_1 ,F_2 } \right\}_{PB} \\
&+ \int\limits_\Sigma {\left( {\left\{ {F_1 ,\Phi_0 (\varphi )} \right\}_{PB}\frac{1}{{\pi (\varphi )}}\left\{ {C_0 (\varphi ),F_2 } \right\}_{PB} - \left( {C_0  \leftrightarrow \Phi_0 } \right)} \right)d^M \varphi } 
\end{split}
\end{equation}
where $\Sigma$ parametrises the $M$-dimensional extended object. It follows that
\begin{equation}
\label{Dirac brackets of canonical variables}
\begin{split}
\left\{ { x^k(\varphi), p_l(\tilde\varphi)} \right\}_{DB} &= \delta^k_{l}\delta(\varphi,\tilde\varphi), \quad k,l=1,\ldots,D-2,\\
\left\{ {\zeta(\varphi),\pi(\tilde\varphi)} \right\}_{DB} &= \delta(\varphi,\tilde\varphi).
\end{split}
\end{equation}
After replacing all Poisson brackets with the Dirac brackets \Ref{Dirac brackets of canonical variables}, the constraints  $C_0$ and $\Phi_0$ can be set strongly to zero. With no risk of confusion, we can drop the subscript $_{DB}$ in \Ref{Dirac brackets of canonical variables} and from now on refer to 
\begin{equation}
\left\{ {F_1,F_2} \right\} := \int\limits_\Sigma  {\left(\frac{\delta F_1}{\delta x^k (\varphi )}\frac{\delta F_2}{\delta p_k (\varphi )} +\frac{\delta F_1}{\delta\zeta(\varphi )}\frac{\delta F_2}{\delta \pi (\varphi )} -\left(F_1\leftrightarrow F_2\right) \right)d^M \varphi}
\end{equation}
as the dynamical Poisson bracket (thus replacing \Ref{Canonical Poisson brackets}). 

The action \Ref{Dirac-Nambu-Goto action} can then be written
\begin{equation}
S = \int \left( {\pi\dot\zeta  + \vec{p}\cdot \dot{\vec{x}}  - \frac{{\vec{p}\,}^2 + g}{-2\pi}}\right) \, d\varphi^0 d^M\varphi 
\end{equation}
with the corresponding Hamiltonian (see also \cite{Hoppe2002})
\begin{equation}
\label{Hamiltonian}
H[\vec{x},\zeta ;\vec{p},\pi] := \int\limits_\Sigma{\frac{{\vec{p}\,}^2 + g}{-2\pi}\, d^M \varphi}, 
\end{equation}
where
\begin{equation}
g = \det(g_{ab}), \quad g_{ab} := \partial_a \vec{x}\cdot\partial_b\vec{x}, \quad a,b=1,\ldots,M,
\end{equation}
and the Euclidean vectors $\vec{x}$ and $\vec{p}$ having 
components $x^k$ and $p_k$, $k=1,\ldots,D-2$, respectively. The remaining first class constraints become
\begin{equation}
\label{Constraints}
C_a={\pi \partial_a\zeta + \vec{p} \cdot \partial_a\vec{x}  \approx 0}, \quad a=1,\ldots,M,
\end{equation}
which, by eliminating $\zeta$, locally give the integrability conditions
\begin{equation}
\label{Integrability criterion}
\partial_a\left(\frac{\vec{p}}{\pi}\right)\cdot \partial_b\vec{x} - \partial_b\left(\frac{\vec{p}}{\pi}\right)\cdot \partial_a\vec{x}  \approx 0, \quad 1\leq a<b\leq M.
\end{equation}
Since the Hamiltonian \Ref{Hamiltonian} does not depend on the phase-space field $\zeta$, its conjugate momentum is conserved, $\dot\pi(\varphi)  = 0$. We can therefore write for the solution to this equation of motion
\begin{equation}
\label{Definition of eta}
\pi(\varphi) = -\eta\rho(\varphi),
\end{equation}
with a discrete dynamical variable $\eta$ satisfying $\dot\eta = 0$, and a non-dynamical positive density $\rho$ normalized such that
\begin{equation}
\label{Normalization of rho}
\int\limits_\Sigma \rho(\varphi)d^M \varphi = 1.
\end{equation}
Furthermore, \Ref{Constraints} allows us to solve for $\zeta$ (in terms of $\vec{x}$, $\vec{p}$ and $\eta$) up to the zero mode
\begin{equation}
\zeta_0 := \int\limits_\Sigma \rho(\varphi) \zeta(\varphi)d^M\varphi.
\end{equation}
The plan of this note is to perform a further partial Hamiltonian reduction 
such that the canonical fields $\zeta$ and $\pi$ can be discarded in favour 
of the discrete zero mode variables $\zeta_0$ and $\eta$, the latter satisfying
\begin{equation}
\label{Zero mode bracket}
\left\{ {\zeta_0,\eta} \right\} = -1,
\end{equation}
and with the field variables $\vec{x}$ and $\vec{p}$ then constrained (locally) by (cf. \Ref{Integrability criterion})
\begin{equation}
\Phi_{ab}:=\partial_a\left(\frac{\vec{p}}{\rho}\right)\cdot \partial_b\vec{x} - \partial_b\left(\frac{\vec{p}}{\rho}\right)\cdot \partial_a\vec{x}  \approx 0, \quad 1\leq a < b \leq M.
\end{equation} 

\section{Constraints}
In this section, we perform a partial gauge fixing corresponding to 
\Ref{Definition of eta}, and introduce the primary constraint that 
becomes second class as a result of this gauge fixing. 

In the following, the domain of integration is always $\Sigma$. 
Introduce a metric $\rho_{ab}$ on $\Sigma$ such that  $\sqrt {\det(\rho_{ab})} = \rho$
and denote the 
non-constant
eigenfunctions, respectively eigenvalues, of the corresponding Laplacian on $\Sigma$ 
by $Y_\alpha$ and $-\mu_\alpha$ respectively, i.e.
\begin{equation}
\label{Eigenvalue equation}
- \Delta Y_\alpha = - \frac{1}{\rho}\partial_a \left( \rho \partial^a \right) Y_\alpha  = \mu_\alpha Y_\alpha, \quad \alpha = 1,2,\ldots
\end{equation}
with $\partial^a:=\rho^{ab} \partial_b$. The eigenfunctions are normalized such that
\begin{equation}
\int{ Y_\alpha(\varphi) Y_\beta(\varphi) \rho(\varphi) d^M\varphi }  = \delta_{\alpha,\beta}
\end{equation}
resulting in 
\begin{equation}
\label{Completeness relation}
\rho(\varphi) + \sum\limits_{\alpha=1}^\infty{\rho(\varphi)Y_\alpha(\varphi)Y_\alpha(\tilde\varphi)} = \delta(\varphi,\tilde\varphi).
\end{equation}
Defining 
\begin{equation}
\label{Green function}
G(\varphi,\tilde\varphi ):=-\sum\limits_{\alpha = 1}^\infty  {\frac{1}{\mu_\alpha}}Y_\alpha(\varphi)Y_\alpha(\tilde\varphi)
\end{equation}
one has \cite{Goldstone}
\begin{equation}
\label{Laplace of Green's function}
\Delta_{\tilde\varphi} G(\varphi,\tilde\varphi) = \frac{\delta(\varphi,\tilde\varphi )}{\rho(\varphi)} - 1
\end{equation}
and
\begin{equation}
\label{Vanishing integral of Green function}
\int{G(\varphi,\tilde\varphi)\rho(\varphi)d^M \varphi} = -\sum\limits_{\alpha = 1}^\infty {\frac{1}{\mu_\alpha}}Y_\alpha(\tilde\varphi) \delta_{0,\alpha} = 0.
\end{equation}
\begin{definition}
Introduce the constraint functions
\begin{equation}
\label{Constraint functions}
\begin{split}
\phi_1 (\varphi ) &:= \zeta (\varphi ) - \int{\zeta (\tilde \varphi )\rho  (\tilde \varphi )d^M \tilde \varphi }  + \Phi (\varphi),\\
\phi_2 (\varphi) &:= \pi(\varphi)-\rho(\varphi)\int{\pi(\tilde\varphi)d^M \tilde\varphi},
\end{split}
\end{equation}
with
\begin{equation}
\label{Auxiliary constraint function}
\Phi (\varphi) := \int{ G^a(\varphi,\tilde\varphi)\frac{\vec p(\tilde \varphi ) \cdot \tilde \partial _a \vec x(\tilde \varphi )}{\pi (\tilde \varphi )}\rho(\tilde\varphi)d^M \tilde \varphi }
\end{equation}
and (introduced in \cite{Goldstone}) $G^a(\varphi,\tilde\varphi) := -\tilde\partial^a G(\varphi,\tilde\varphi)$.
\end{definition}
Using the primary constraints \Ref{Constraints} in \Ref{Auxiliary constraint function}, integrating by parts, and applying \Ref{Laplace of Green's function} gives
\begin{equation}
\begin{split}
\phi_1 (\varphi ) &\approx \zeta (\varphi ) - \int {\zeta (\tilde \varphi )\rho (\tilde \varphi )d^M \tilde \varphi }  + \int {G^a (\varphi ,\tilde \varphi )\left( { - \tilde \partial _a \zeta (\tilde \varphi )} \right)\rho (\tilde \varphi )d^M \tilde \varphi } \\
&= \zeta (\varphi ) - \int {\zeta (\tilde \varphi )\rho  (\tilde \varphi )d^M \tilde \varphi }  - \int {\zeta (\tilde \varphi )\rho (\tilde \varphi )\Delta _{\tilde \varphi } G(\varphi,\tilde\varphi )d^M \tilde \varphi } =0,
\end{split}
\end{equation}
which means that $\phi_1$ is a primary constraint. Indeed, one easily checks that
\begin{equation}
\phi _1 (\varphi ) = \int {G^a (\varphi ,\tilde \varphi )\frac{1}{{\pi (\tilde \varphi )}}C_a (\tilde \varphi )\rho (\tilde \varphi )d^M \tilde \varphi }
\end{equation}
with $C_a$ given in \Ref{Constraints}. We now impose
\begin{equation}
\label{Gauge constraint}
\phi_2 (\varphi) \approx 0
\end{equation}
as an additional gauge constraint. Using this, one can solve for $\pi$ in terms of the discrete variable $\eta = - \int \pi d^M\varphi$ (cf. \Ref{Definition of eta}) on the constraint surface. 
\begin{lemma}
\label{Lemma Poisson brackets of constraint functions}
The constraints \Ref{Constraint functions} satisfy the dynamical Poisson brackets
\begin{equation}
\label{Poisson brackets of constraint functions 1-1}
\left\{ {\phi_1 (\varphi ),\phi_1 (\tilde\varphi)} \right\} \approx 0,
\end{equation}
\begin{equation}
\label{Poisson brackets of constraint functions 2-2}
\left\{ {\phi_2 (\varphi ),\phi_2 (\tilde\varphi)} \right\} = 0,
\end{equation}
and
\begin{equation}
\label{Poisson brackets of constraint functions 1-2}
\left\{ {\phi_1 (\varphi),\phi_2 (\tilde \varphi )} \right\} = \delta (\varphi, \tilde \varphi ) - \rho (\tilde \varphi ).
\end{equation}
\begin{proof}
The following variational derivatives of \Ref{Auxiliary constraint function} will be needed:
\begin{equation}
\label{Variations of Auxiliary constraint function}
\begin{split}
\frac{{\delta \Phi(\varphi)}}{{\delta \zeta (\tilde \varphi )}} &= 0,\\
\frac{\delta \Phi(\varphi)}{\delta \pi (\tilde \varphi )} &= -G^a(\varphi,\tilde\varphi)\frac{{\vec p(\tilde \varphi ) \cdot \tilde \partial_a \vec x(\tilde \varphi )}}{{{\pi(\tilde\varphi )}^2 }}\rho(\tilde\varphi),\\
\frac{{\delta \Phi(\varphi)}}{{\delta x^k (\tilde \varphi )}} &=  - \tilde \partial _a \left( {G^a(\varphi ,\tilde \varphi )}\frac{{p_k (\tilde \varphi )}}{{\pi (\tilde \varphi )}}\rho (\tilde \varphi ) \right),\\
\frac{\delta \Phi(\varphi)}{\delta p_k (\tilde \varphi )} &=G^a(\varphi,\tilde\varphi)\frac{\tilde \partial _a x^k (\tilde \varphi )}{{\pi (\tilde \varphi )}}\rho (\tilde \varphi ).
\end{split}
\end{equation}  
Inserting \Ref{Constraint functions} on the left hand side of \Ref{Poisson brackets of constraint functions 1-1} yields
\begin{equation*}
\left\{ {\phi_1 (\varphi ),\phi_1 (\tilde \varphi )} \right\} = \frac{{\delta \Phi (\tilde \varphi )}}{{\delta \pi (\varphi )}}
- \frac{{\delta \Phi (\varphi )}}{{\delta \pi (\tilde\varphi )}} + \int {\left(\frac{{\delta \Phi (\varphi )}}{{\delta \pi (\hat\varphi )}} -\frac{{\delta \Phi (\tilde\varphi )}}{{\delta \pi (\hat\varphi )}}\right) \rho(\hat\varphi)d^M \hat \varphi } + \left\{ {\Phi (\varphi ),\Phi (\tilde \varphi )} \right\}
\end{equation*}
where we have used that
\begin{equation*}
\left\{ {\zeta (\varphi ),\Phi(\tilde\varphi)} \right\} = \frac{{\delta \Phi(\tilde\varphi)}}{{\delta \pi (\varphi )}}.
\end{equation*}
Using the definition of the Poisson bracket, the variational derivatives \Ref{Variations of Auxiliary constraint function}, and integrating by parts gives
\begin{equation*}
\begin{split}
\left\{ {\Phi (\varphi ),\Phi (\tilde \varphi )} \right\} &= \int {\frac{{\vec{p}(\hat \varphi ) \cdot \hat \partial _b \vec{x}(\hat \varphi )}}{{\pi (\hat \varphi )^2 }}\rho (\hat \varphi )^2 \left( {G^b (\tilde \varphi ,\hat \varphi )\Delta _{\hat \varphi } G(\varphi ,\hat \varphi ) - G^b (\varphi ,\hat \varphi )\Delta _{\hat \varphi } G(\tilde \varphi ,\hat \varphi )} \right)d^M \hat \varphi }\\
&- \int {G^a (\varphi ,\hat \varphi )G^b (\tilde \varphi ,\hat \varphi )\frac{{\rho (\hat \varphi )^2 }}{{\pi (\hat \varphi )}}\left( {\hat \partial _a \left( {\frac{{\vec{p}}}{{\pi }}} \right) \cdot \hat \partial _b \vec{x} - \hat \partial _b \left( {\frac{{\vec{p}}}{{\pi}}} \right) \cdot\hat \partial _a \vec{x}} \right)d^M \hat \varphi } .
\end{split}
\end{equation*}
Applying \Ref{Integrability criterion} to the expression within parenthesis in the last line above, together with the use of equations \Ref{Laplace of Green's function} and \Ref{Variations of Auxiliary constraint function}, leads to
\begin{equation*}
\left\{ {\Phi (\varphi ),\Phi (\tilde \varphi )} \right\} \approx  - \frac{{\delta \Phi (\tilde \varphi )}}{{\delta \pi (\varphi )}} + \frac{{\delta \Phi (\varphi )}}{{\delta \pi (\tilde \varphi )}} - \int {\left( {\frac{{\delta \Phi (\varphi )}}{{\delta \pi (\hat \varphi )}} - \frac{{\delta \Phi (\tilde \varphi )}}{{\delta \pi (\hat \varphi )}}} \right)\rho (\hat \varphi )d^M \hat \varphi }, 
\end{equation*}
which proves \Ref{Poisson brackets of constraint functions 1-1}. The Poisson bracket \Ref{Poisson brackets of constraint functions 2-2} is trivial since $\phi_2 (\varphi)$ only involves the field $\pi(\varphi)$. Finally, \Ref{Poisson brackets of constraint functions 1-2} is easily proven using \Ref{Canonical Poisson brackets}, \Ref{Normalization of rho}, and by noting that 
\begin{equation*}
\left\{ {\Phi(\varphi ),\pi (\tilde\varphi )} \right\}=0,
\end{equation*}
since $\Phi$ does not contain $\zeta$.
\end{proof}
\end{lemma}

\section{Dirac bracket}
In this section, we define a Dirac bracket that turns the second class constraints \Ref{Constraint functions} effectively first class. To this end, the canonical procedure would be to set
\begin{equation}
\label{Constraint matrix}
C_{ab}(\varphi,\tilde\varphi):= \left\{ {\phi_a(\varphi),\phi_b(\tilde\varphi)} \right\}, \quad a,b=1,2,
\end{equation}
and then define the Dirac bracket by
\begin{equation}
\label{Dirac bracket}
\left\{{F_1,F_2}\right\}^*=\left\{{F_1,F_2}\right\} - \sum_{a,b=1,2}\int\left\{{F_1,\phi_a(\varphi)}\right\} (C^{-1})_{ab}(\varphi,\tilde\varphi) \left\{{\phi_b(\tilde\varphi),F_2}\right\}d^M\varphi d^M\tilde\varphi.
\end{equation}
However, as it stands, this expression is not well defined since \Ref{Constraint matrix} is not invertible. More specifically, by equation \Ref{Poisson brackets of constraint functions 1-2} in Lemma \ref{Lemma Poisson brackets of constraint functions}, one finds that
\begin{equation}
\label{Chi}
\chi(\varphi,\tilde\varphi) := C_{12}(\varphi,\tilde\varphi)
\end{equation}
defines a projection onto the space of zero-mean real-valued functions on $\Sigma$:
\begin{equation}
\left[ {\chi f} \right](\varphi ) := \int {\chi (\varphi ,\tilde \varphi )f(\tilde \varphi )d^M \tilde \varphi }  = f(\varphi ) - \bar f
\end{equation}
with the mean of $f$ defined by
\begin{equation}
\bar f := \int {f(\varphi)\rho (\varphi)d^M \varphi }.
\end{equation}
It follows that $\ker \left( \chi \right) = \left\{ {f:\Sigma \to \mathbb{R}\,|\,f {\text{ constant}}} \right\}$ is non-trivial, and \Ref{Constraint matrix} is indeed non-invertible. 
Fortunately, it turns out that the bracket obtained by restricting \Ref{Chi} to $\text{im}\left( \chi \right) = \left\{ {f:\Sigma \to \mathbb{R}\,|\,\bar f = 0} \right\}$, 
i.e. $\left.\chi \right|_{\text{im}\left( \chi \right)}=\delta (\varphi,\tilde\varphi)$, 
will be sufficient for our purposes.
\begin{definition}
\label{Definition Dirac bracket}
Define
\begin{equation}
\label{Alternative Dirac bracket}
\left\{ {F_1,F_2} \right\}^*  := \left\{ {F_1,F_2} \right\} + \int {\Bigl( {\left\{ {F_1,\phi _1 (\varphi )} \right\}\left\{ {\phi _2 (\varphi),F_2} \right\} - \left\{ {F_1,\phi _2 (\varphi )} \right\}\left\{ {\phi _1 (\varphi ),F_2} \right\}} \Bigr)d^M \varphi }.
\end{equation}
\end{definition}
This bracket has the desired properties, as seen by the following:
\begin{lemma}
The bracket \Ref{Alternative Dirac bracket} satisfies
\begin{equation}
\label{Second class requirement}
\left\{ {\phi_k,\phi_l} \right\}^*  \approx 0, \quad k,l=1,2.
\end{equation}
Furthermore, for arbitrary functions of $\vec{x}$, $\zeta_0$, $\vec{p}$, 
and $\pi$ (in particular $\eta$), it holds that
\begin{equation}
\label{Antisymmetry}
\left\{ {F_1,F_2} \right\}^*  = -\left\{ {F_2,F_1} \right\}^*,
\end{equation}
\begin{equation}
\label{Product law}
\left\{ {F_1,F_2F_3} \right\}^*  = \left\{ {F_1,F_2} \right\}^*F_3 + F_2\left\{ {F_1,F_3} \right\}^*,
\end{equation}
\begin{equation}
\label{Jacobi identity}
\left\{ {\left\{ {F_1,F_2} \right\}^* ,F_3} \right\}^*  + \left\{ {\left\{ {F_2,F_3} \right\}^* ,F_1} \right\}^* +\left\{ {\left\{ {F_3,F_1} \right\}^* ,F_2} \right\}^* = 0,
\end{equation}
and
\begin{equation}
\label{Bracket between constraint and F}
\left\{F,{\phi_k} \right\}^*  = 0, \quad k=1,2.
\end{equation}
\begin{proof}
The brackets \Ref{Second class requirement} follow immediately using Lemma \ref{Lemma Poisson brackets of constraint functions} and the projection property of $\chi$, while relations \Ref{Antisymmetry} and \Ref{Product law} are trivial. The Jacobi identity is proven along the same lines as the original proof by Dirac \cite{Dirac1950}. To simplify the algebra, we first write \Ref{Alternative Dirac bracket} symbolically as
\begin{equation*}
\begin{split}
\left\{ {F_1 ,F_2 } \right\}^*  &= \left\{ {F_1 ,F_2 } \right\} + \int {\int {\left\{ {F_1 ,\phi _a (\varphi )} \right\}M_{ab} (\varphi ,\tilde \varphi )\left\{ {\phi _b (\tilde \varphi ),F_2 } \right\}d^M\varphi d^M \tilde \varphi } }  \\ 
&=\left\{ {F_1 ,F_2 } \right\} + \left\{ {F_1 ,\phi _a } \right\}M_{ab} \left\{ {\phi _b ,F_2 } \right\} 
\end{split}
\end{equation*}
with $M_{11}=M_{22}=0$ and $M_{12}(\varphi ,\tilde \varphi )=-M_{21}(\varphi ,\tilde \varphi )=\delta(\varphi , \tilde \varphi )$. Let $\mathcal{S}$ denote cyclic permutation of 1, 2, 3, and summation of the result, such that
\begin{equation*}
\mathcal{S} \left(\left\{ {\left\{ {F_1,F_2} \right\}^* ,F_3} \right\}^*\right)  = \left\{ {\left\{ {F_1,F_2} \right\}^* ,F_3} \right\}^*  + \left\{ {\left\{ {F_2,F_3} \right\}^* ,F_1} \right\}^* +\left\{ {\left\{ {F_3,F_1} \right\}^* ,F_2} \right\}^*.
\end{equation*}
Then
\begin{equation*}
\begin{split}
\mathcal{S} \left(\left\{ {\left\{ {F_1 ,F_2 } \right\}^* ,F_3 } \right\}^* \right) &= \mathcal{S}\left(\left\{ {\left\{ {F_1 ,F_2 } \right\},F_3 } \right\}\right) \\ 
&+ M_{rs} \mathcal{S}\left( {\left\{ {F_1 ,\phi _r } \right\}\left( {\left\{ {\left\{ {\phi _s ,F_2 } \right\},F_3 } \right\} + \left\{ {\left\{ {F_3 ,\phi _s } \right\},F_2 } \right\} + \left\{ {\left\{ {F_2 ,F_3 } \right\},\phi _s } \right\}} \right)} \right) \\ 
&+ M_{rs} M_{tu} \mathcal{S}\left( {\left\{ {\left\{ {F_1 ,\phi _r } \right\}\left\{ {\phi _s ,F_2 } \right\},\phi _t } \right\}\left\{ {\phi _u ,F_3 } \right\}} \right)
\end{split}
\end{equation*}
The first two lines are zero by the Jacobi identity for the Poisson bracket. Renaming dummy summation indices and again using the Jacobi identity, the last line can be rewritten as
\begin{equation*}
\begin{split}
\mathcal{S} \left(\left\{ {\left\{ {F_1 ,F_2 } \right\}^* ,F_3 } \right\}^*\right)  &= M_{rs} M_{tu} \mathcal{S}\left( {\left\{ {\phi _s ,F_2 } \right\}\left\{ {\phi _u ,F_3 } \right\}\left( {\left\{ {\left\{ {F_1 ,\phi _r } \right\},\phi _t } \right\} + \left\{ {\left\{ {\phi _t ,F_1 } \right\},\phi _r } \right\}} \right)} \right) \\ 
&=  - M_{rs} M_{tu} \mathcal{S}\left( {\left\{ {\phi _s ,F_2 } \right\}\left\{ {\phi _u ,F_3 } \right\}\left\{ {\left\{ {\phi _r ,\phi _t } \right\},F_1 } \right\}} \right)
\end{split}
\end{equation*}
By Lemma \ref{Lemma Poisson brackets of constraint functions}, this simplifies to
\begin{equation*}
\mathcal{S} \left(\left\{ {\left\{ {F_1 ,F_2 } \right\}^* ,F_3 } \right\}^*\right) =  - M_{12} M_{12} \mathcal{S}\left( {\left\{ {\phi _2 ,F_2 } \right\}\left\{ {\phi _2 ,F_3 } \right\}\left\{ {\left\{ {\phi _1 ,\phi _1 } \right\},F_1 } \right\}} \right)
\end{equation*}
when summing over indices. But this is zero since 
\begin{equation*}
{\left\{ {\zeta _0 ,\phi _2 (\varphi )} \right\} = 0} \quad \Rightarrow \quad {\left\{ {F ,\phi _2 (\varphi )} \right\} = 0}
\end{equation*}
for functions $F$ of $\vec{x}$, $\zeta_0$, $\vec{p}$, and $\pi$. To prove \Ref{Bracket between constraint and F}, we first note that 
\begin{equation*}
\begin{split}
\int {\left\{ {F,\phi_1 (\varphi )} \right\}\rho (\varphi )d^M \varphi }  &= 
\int {\left( {\frac{{\delta F}}{{\delta x^k (\tilde \varphi )}}\frac{{\delta \Phi (\varphi )}}{{\delta p_k (\tilde \varphi )}} + \frac{{\delta F}}{{\delta \zeta (\tilde \varphi )}}\frac{{\delta \Phi (\varphi )}}{{\delta \pi (\tilde \varphi )}}} \right.}  \\ 
&\left. { - \frac{{\delta \Phi (\varphi )}}{{\delta x^k (\tilde \varphi )}}\frac{{\delta F}}{{\delta p_k (\tilde \varphi )}} - \frac{{\delta \phi _1(\varphi) }}{{\delta \zeta (\tilde \varphi )}}\frac{{\delta F}}{{\delta \pi (\tilde \varphi )}}} \right)\rho (\varphi )d^M \varphi d^M \tilde \varphi\\
&= - \int {\frac{{\delta \phi _1(\varphi ) }}{{\delta \zeta (\tilde \varphi )}}\frac{{\delta F}}{{\delta \pi (\tilde \varphi )}}\rho (\varphi )d^M \varphi d^M \tilde \varphi }\\
&=- \int {\left( {\delta (\varphi, \tilde \varphi ) - \rho (\tilde \varphi )} \right)\frac{{\delta F}}{{\delta \pi (\tilde \varphi )}}\rho (\varphi )d^M \varphi d^M \tilde \varphi }=0
\end{split}
\end{equation*}
where, in the second equality, we have used that variations of $\Phi$ \Ref{Variations of Auxiliary constraint function} depend explicitly on $G^a(\varphi,\tilde\varphi)$. Analogous to \Ref{Vanishing integral of Green function}, all these terms vanish when one integrates with respect to $\varphi$ above. It follows that for an arbitrary function $F$ of $\vec{x}$, $\zeta_0$, $\vec{p}$, and $\pi$,
\begin{equation*}
\begin{split}
\left\{ {F,\phi _1 (\varphi )} \right\}^*  &= \int {\bigl( {\left\{ {F,\phi _1 (\hat \varphi )} \right\}\rho (\hat \varphi ) - \left\{ {F,\phi _2 (\hat \varphi )} \right\}\left\{ {\phi _1 (\hat \varphi ),\phi _1 (\varphi )} \right\}} \bigr)d^M \hat \varphi }  = 0,\\
\left\{ {F,\phi _2 (\varphi )} \right\}^*  &= \rho (\varphi )\int {\left\{ {F,\phi _2 (\hat \varphi )} \right\}d^M \hat \varphi }  = 0.
\end{split}
\end{equation*}
\end{proof}
\end{lemma}
The Hamiltonian reduction of the degrees of freedom is completed by:
\begin{theorem} \label{Reduction Theorem}
The phase space variables $\vec{x}$, $\zeta_0$, $\vec{p}$, and $\eta$ satisfy
\begin{equation}
\{\zeta_0,\eta\}^*=-1, \quad
\left\{ { x^k(\varphi), p_l(\tilde\varphi)} \right\}^* = \delta^k_{l}\delta(\varphi,\tilde\varphi), \quad k,l=1,\ldots D-2,
\end{equation}
with all other brackets zero,
and the reduced Hamiltonian is
\begin{equation}
	H[\vec{x},\zeta_0 ;\vec{p},\eta] = \frac{1}{2\eta} \int\limits_\Sigma{\frac{{\vec{p}\,}^2 + g}{\rho}\, d^M \varphi}. 
\end{equation}
\begin{proof}
We apply Definition \ref{Definition Dirac bracket} (suppressing the arguments of the field variables whenever there is no risk of confusion). Since $\left\{ {\phi_2,x^k} \right\} =\left\{ {\phi_2,p_l} \right\} = 0$, the Dirac brackets between the $\vec{x}$ and $\vec{p}$ coincide with the corresponding dynamical Poisson brackets ($\left\{ {x^k,x^l} \right\}^* = \left\{ {x^k,x^l} \right\}$, etc.). Furthermore, 
\begin{equation*}
\begin{split}
\left\{ {\zeta _0 ,\eta } \right\}^*  &= \left\{ {\zeta _0 ,\eta } \right\} + \int {\left( {\left\{ {\zeta _0 ,\phi_1 (\varphi )} \right\}\left\{ {\phi_2 (\varphi ),\eta } \right\} - \left\{ {\zeta _0 ,\phi_2 (\varphi )} \right\}\left\{ {\phi_1 (\varphi ),\eta } \right\}} \right)d^M \varphi },  \\ 
\left\{ {\zeta _0 ,x^k} \right\}^*  &= \left\{ {\zeta _0 ,x^k} \right\} - \int {\left\{ {\zeta _0 ,\phi_2 (\varphi )} \right\}\left\{ {\phi_1 (\varphi ),x^k} \right\}d^M\varphi },  \\ 
\left\{ {\zeta _0 ,p_k} \right\}^*  &= \left\{ {\zeta _0 ,p_k} \right\} - \int {\left\{ {\zeta _0 ,\phi_2 (\varphi )} \right\}\left\{ {\phi_1 (\varphi ),p_k} \right\}d^M\varphi },  \\ 
 \left\{ {\eta ,x^k} \right\}^*  &= \left\{ {\eta ,x^k} \right\} - \int {\left\{ {\eta ,\phi_2 (\varphi )} \right\}\left\{ {\phi_1 (\varphi ),x^k} \right\}d^M\varphi },  \\ 
 \left\{ {\eta ,p_k} \right\}^*  &= \left\{ {\eta ,p_k} \right\} - \int {\left\{ {\eta ,\phi_2 (\varphi )} \right\}\left\{ {\phi_1 (\varphi ),p_k} \right\}d^M\varphi }. 
\end{split}
\end{equation*}
The result now follows by using \Ref{Canonical Poisson brackets}, \Ref{Zero mode bracket}, and the fact that the dynamical Poisson brackets 
\begin{equation*}
\left\{ {\zeta _0 ,\phi_1} \right\},\quad \left\{ {\zeta _0 ,\phi_2} \right\},\quad \left\{ {\zeta _0 ,x^k} \right\},\quad \left\{ {\zeta _0 ,p_k} \right\},\quad \left\{ {\eta ,\phi_1} \right\},\quad 
\left\{ {\eta ,\phi_2} \right\},\quad \left\{ {\eta ,x^k} \right\},\quad \left\{ {\eta ,p_k} \right\},
\end{equation*}
all vanish.
\end{proof}
\end{theorem}

\subsection*{Acknowledgments} 
This work was supported by the G\"oran Gustafsson Foundation 
and the Knut and Alice Wallenberg Foundation (grant KAW 2005.0098).

\appendix

\section{Other notational conventions}
\label{Appendix Other conventions}
For the convenience of the reader, some relations between our notation and another one often used in the literature are summarised in this appendix.
Taking the metric signature $\eta \sim (+,-,\ldots,-)$, we have followed here the convention that light-cone coordinates are defined by
$$
	x^+ := \frac{1}{2}(x^0 + x^{D-1}),
	\qquad
	x^- \equiv \zeta := x^0 - x^{D-1},
$$
and the scalar product is given by
$$
	a \cdot b = a^\mu b_\mu = a^+ b^- + a^- b^+ - \vec{a} \cdot \vec{b}.
$$
On the other hand, it is also common in the literature to use reversed metric signature
$\tilde{\eta} \sim (-,+,\ldots,+)$, with the convention
$$
	\tilde{x}^{\pm} := \frac{1}{\sqrt{2}}(x^{D-1} \pm x^0),
$$
and scalar product
$$
	\tilde{a} \cdot \tilde{b} 
	= \tilde{a}^\mu \tilde{b}_\mu 
	= \tilde{a}^+ \tilde{b}^- + \tilde{a}^- \tilde{b}^+ + \tilde{\vec{a}} \cdot \tilde{\vec{b}} 
	= -a \cdot b,
$$
so that a translation between notations is given by
$$
	\tilde{x}^+ = \sqrt{2} \, x^+,
	\qquad
	\tilde{x}^- = -\frac{1}{\sqrt{2}} \zeta,
	\qquad
	\tilde{\vec{x}} = \vec{x}.
$$
After gauge fixing $\varphi^0 \stackrel{!}{=} \tilde{x}^+$, the conjugate momenta become (the expression for $\tilde{p}^-$ following from the constraint $C_0$ in \Ref{All primary constraints})
$$
	\tilde{p}^+ := \frac{\partial \mathcal{L}}{\partial (\partial_0 \tilde{x}^-)}
	= -\sqrt{2} \, \pi,
	\qquad
	\tilde{p}^- = -\frac{{\tilde{\vec{p}} \,}^2 + g}{2\tilde{p}^+},
	\qquad	
	\tilde{\vec{p}} = \vec{p},
$$
and the light-cone Hamiltonian
$$
	\tilde{H} = -\tilde{P}^- = \frac{1}{\sqrt{2}} \, H,
	\qquad
	\tilde{P}^- := \int_\Sigma \tilde{p}^- \,d^M\varphi.
$$
The zero mode variables remaining after the reduction of the phase space are
$$
	\tilde{X}^- := \int_\Sigma \tilde{x}^- \rho\,d^M\varphi = -\frac{1}{\sqrt{2}} \zeta_0,
	\qquad
	\tilde{P}^+ := \int_\Sigma \tilde{p}^+ \,d^M\varphi= \sqrt{2} \, \eta,
$$
with $\{ \tilde{X}^-, \tilde{P}^+ \}^* = - \{ \zeta_0, \eta \}^* = 1$. The mass-squared of the theory is given by
$$
	\mathbb{M}^2 := P \cdot P = -\tilde{P} \cdot \tilde{P}
	= -2\tilde{P}^+ \tilde{P}^- - {\tilde{\vec{P}} \,}^2
	= \int_\Sigma \frac{{\vec{p} \,}^2 + g}{\rho}\,d^M\varphi 
	- \left( \int_\Sigma \vec{p}\,d^M\varphi \right)^2.
$$

\end{document}